\documentclass[journal,11pt,twocolumn]{IEEEtran}
\usepackage{amsmath}
\usepackage{amsfonts}
\usepackage{amsthm}
\usepackage{dsfont}
\usepackage{epsfig}
\usepackage{upgreek}
\usepackage{amssymb}
\usepackage{graphicx}
\usepackage{setspace}

\theoremstyle{definition}

\newcommand{\RowCS}{\mbox{\small\textregistered}}
\newcommand{\ColCS}{\mbox{\small\copyright}}

\begin{document}

\title{{Software Adaptation and Generalization of \\
        Physically-Constrained Games} }       
\author{
%{\large\bf Jeffrey Uhlmann}\\
%University of Missouri-Columbia}
\IEEEauthorblockN{{\large\bf Jeffrey Uhlmann}}\\
\IEEEauthorblockA{\small University of Missouri-Columbia\\
201 Naka Hall, Columbia, MO 65211\\
Email: uhlmannj@missouri.edu}}
\date{}          
\maketitle
\thispagestyle{empty}

%For singlespaced version
%\vspace{-11pt}
%For doublespaced version
\vspace{-0.5in}

\begin{abstract}
In this paper we provide a case study of the use of relatively 
sophisticated mathematics and algorithms to redefine and adapt
a simple traditional game/puzzle to exploit the computational
power of smart devices. The focus here is not so much on the end
product as it is on the process and considerations underpinning
its development. Ancillary results of the venture include 
generalizations of the circular-shift operator and examination 
of its computational complexity. 
~\vspace{3pt} \\ 
\begin{footnotesize}
\end{footnotesize}
\end{abstract}

\section{Introduction}

The subject of ``game design'' is far too broad to admit any form of concise top-down treatment, 
but often it is possible to glean general concepts and principles from bottom-up consideration of
concrete examples. In the case of particular board games, e.g., {\em Monopoly}, it is straightforward to 
identify specific features that people find to be engaging and entertaining\,\cite{board,monopoly}, 
and from these it's possible to identify general principles that explain the appeal 
not only of similar board games but also a wider
variety of other games as well. However, these general principles would not be sufficient to 
permit an alien scholar to predict, or derive as being inevitable, the actual existence of such games 
{\em as a genre}. That's because the genre's existence has less to do with objectively
discernible attributes and more to do with historical happenstance: someone developed the
first instance of the kind, it became popular, and subsequent similar games were developed to 
leverage not only the popular attributes of that first game but also the newly existing 
{\em familiarity} of those attributes among general consumers. In other words, 
once the initial instance became popular the genre could evolve incrementally with 
improvements that would be appreciated by consumers without the steep learning 
curve of a completely {\em de novo} game. 

The availability of relatively low-cost computers in the 1970s created opportunities
to not only design a new class of games tailored to leverage the unique capabilities
of the new technology but also to adapt existing games for computer implementation.
A good example is the popular card game {\em Solitaire}\footnote{This term
is actually generic for a variety of single-player card games (also known as 
{\em Patience}) but is also commonly used to refer to the specific
game {\em Klondike}\,\cite{sol1,sol2}. Because more people know that specific game
under the name ``Solitaire'' instead of ``Klondike,'' we will use it in that sense
here.}. Variations of this game are found pre-installed on most
home computers and smart devices. It is interesting to note which features
of the game are simulated and which are not. In theory, it would seem that
the game could be implemented using any set of distinct shapes and/or color
designs instead of simulating the suits of a traditional 52-card deck of playing
cards. But to do so would sacrifice the familiarity of the game to consumers.
In other words, while the logic of the game would be identical {\em it
wouldn't be Solitaire} from the perspective of most consumers. 
In fact, current versions of Solitaire differ little from early computer 
implementations even though advances in computing power could permit
a highly realistic simulation of human hands holding and manipulating cards
to more closely emulate how the game is played with real cards. 
This suggests that current implementations include the salient
features necessary to capture the experience and serve as
an acceptable replacement for the traditional form of the game.

Traditional Solitaire (Klondike) emerged as one of the most popular 
and widely-played computer games, and it spawned a family of 
Solitaire-{\em like} games that also became popular despite the
fact that they were not convenient to play with actual cards.
One such example is {\em TriPeaks}\cite{tripeaks}, which requires 
cards to be arranged in three overlapping triangular configurations 
that would be cumbersome to maintain in a neat-and-orderly
form using real cards on a real table. As such, it represents 
one of the first examples of a virtual card game designed
specifically to be played on a computer. Now there are many
other games that are similarly designed to exploit the 
familiarity of Solitaire in a form that leverages the advantages
of computer simulation. In this paper we will examine the
process of adapting and redefining an existing game for
computer simulation in a way that reveals a sequence of
concrete steps in which mathematical and algorithmic
considerations can come into play.

\section{Sliding-Tile Puzzle Games}

Before the prevelance of hand-held computer games there
were many popular puzzle games requiring manual movement
of objects to achieve a goal configuration. Some of these involved 
the movement of wooden pegs within an arrangement of holes in 
a wooden base while others involved the sliding of tiles within a
rectangular frame\,\cite{sliding}. 
An example of the latter is the 8-tile puzzle:
\vspace{-4pt}\begin{center}
\includegraphics[width=0.485\textwidth]{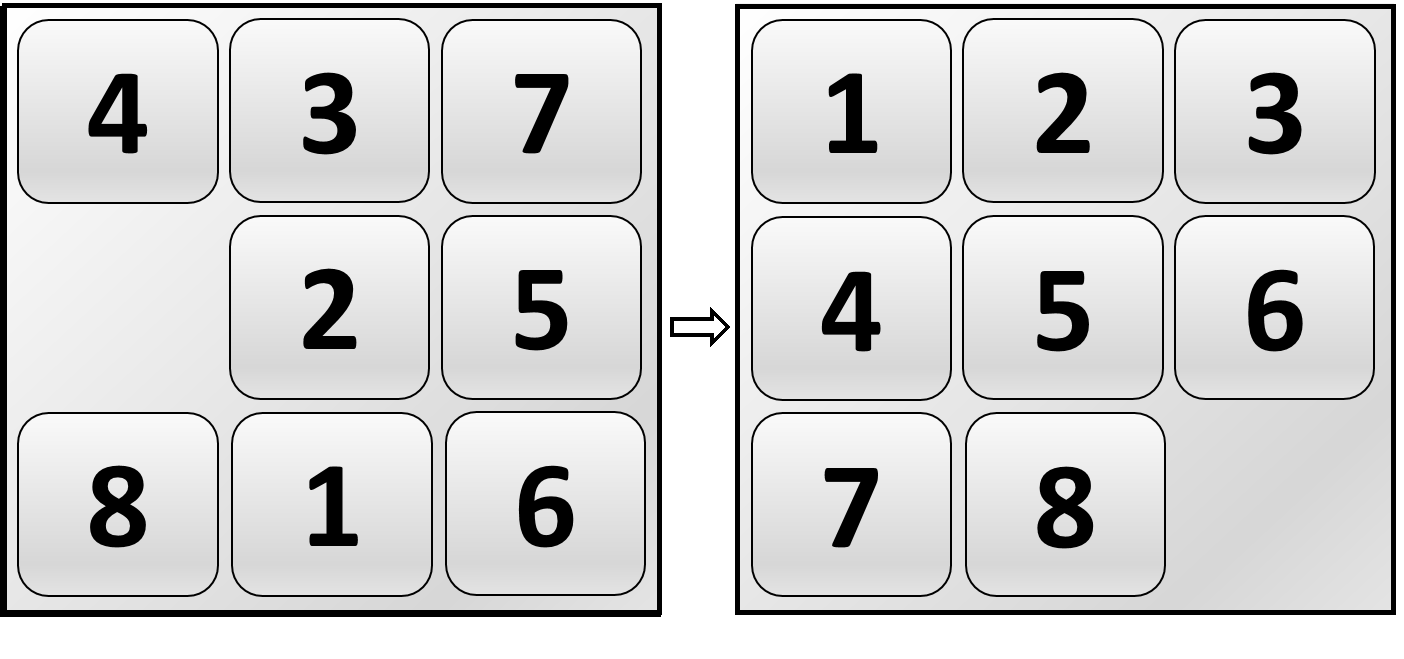}
\end{center}\vspace{-12pt}
where a chosen tile adjacent to the empty space can be moved into
that space\footnote{An alternative, though equivalent, interpretation
of a move is that the space (or {\em hole}) swaps location with the
tile\,\cite{physrev}.} in such a way that a sequence of moves transforms
the random initial configuration ({\em left}) to the ordered goal
configuration ({\em right}). Traditionally the frame and tiles were
made of wood, then later plastic, and the number of tiles could
be larger, e.g., $15$ in a $4\times 4$ frame, and the numeric 
digits might be replaced with letters or other symbols, or even
graphical elements forming an image when placed in the
goal configuration.

Of course a direct simulation of the game is straightforward,
and the above figure provides a reasonable prototype of a
possible visual interface. If one starts with such a direct
translation of the real-world interface then the mechanics
of the interaction might include touch sensitivity with 
velocity detection for the user to specify a particular
tile and direction of motion using a finger swipe. 
On further consideration, however, it can be recognized 
that simple touch detection is sufficient because the 
direction of motion can be uniquely inferred based on 
the location of the space. 

An important question that arises is whether something
critical to the ``game experience'' is lost if swiping is 
replaced with touch\footnote{Similarly, is it worthwhile
to associate sound with the movement of tiles, e.g.,
as they slide and then ``click'' on impact? Sound 
can contribute immensely to user experience 
when playing a simulated physical-based game
like this.}. Fortunately in this case nothing is 
lost because touch detection will be triggered by a 
swipe to produce the same result, i.e., the user
experience will be identical in most circumstances.
At this point we can see how the traditional game
can easily be implemented in a way that preserves
an exact analogy between the real version and
the simulated one. With this we can begin to 
consider ways in which the simulated version can
go beyond what is possible with the real version
due to the physical constraints of tiles on a fixed
frame.

Simulation offers myriad opportunities for changing
the shape of the frame and the shape of the tiles, 
e.g., to triangles or hexagons, because there is 
no need to support the physical sliding of pieces.
Thinking even more broadly, it can be recognized
that in the simulated version there is no need for
an empty space/hole. More specifically, we could 
have nine tiles and define a new operation in
which the swiping of a tile causes it to swap
position with the adjacent tile in the direction
of the swipe. Unfortunately, such an operation
causes the game play to degenerate into
simply executing a trivial sequence of swaps to
move tile $1$ directly to its correct final position,
then doing the same for tile $2$, and so on.

An alternative operation can also be defined
that is in some ways more consistent with 
the spirit of the traditional game. 
That is to allow a swipe to {\em circularly
shift} a given row or column. For example,
if a right swipe is applied to a row with 
digits 4-3-7, the result would be for each
tile to move one position to its right with
the rightmost tile moving (wrapping around)
to the first position, so 4-3-7 would circularly
shift to become 7-4-3. This operation
preserves many of the familiar aspects
of the traditional game in that later
moves become increasingly constrained
as more tiles become fixed in their final
positions. More specifically, the effectiveness
of a simple greedy strategy is limited because
it may move a particular piece closer
to its destination while simultaneously
moving other pieces away from their goal
positions. 

The circular-shift operation provides an
interesting variation on the game, but can
the idea be extended further? For example, 
instead of discrete integer-value shifts can the
operation be generalized to allow a continuous
range of motion, e.g., a circular shift of
$k=1.83$ or $k=-3.24$, as determined 
by the duration of the player's swipe? 
It isn't immediately clear what such a 
generalization would {\em look like}, or
whether it would contribute positively to the
game-playing experience, but it's certainly
worth investigating because it's a feature that 
can potentially be supported on a computer 
but not by any simple physical device. This provides 
an opportunity to examine the circular-shift 
operator in a more abstract mathematical
sense in order to assess whether it is 
possible to generalize it to, e.g., take on arbitrary 
real values.

\section{Generalized Circular-Shift (GCS) Operator}

The {\em shift} operator is widely used in computer engineering and computer science to transform the 
state of a computer register, which can be treated abstractly as a vector of length $n$. The shift operator 
takes an integer parameter $k$ and moves each value at location $i$ of an $n$-element vector to location $i+k$, 
with each of the first $k$ elements becoming zeros. The following is an example with $k=2$:
\begin{center}
\begin{tabular}{|c|c|c|c|c|}
\hline\rule{-2.5pt}{2.5ex}  
1 & 2 & 3 & 4 & 5\\
\hline
\end{tabular}
~ $\rightarrow$ ~
\begin{tabular}{|c|c|c|c|c|}
\hline\rule{-2.5pt}{2.5ex}
0 & 0 & 1 & 2 & 3\\
\hline
\end{tabular}
\end{center}
%\vspace{11pt}

The {\em circular}-shift operator (or circular {\em buffer}\,\cite{circbuff})
is also familiar in engineering applications and is defined analogously except that values 
moved beyond the index range of the vector are moved (rotated) to locations modulo the length of the vector in the 
intuitively natural way we've already assumed:

\begin{center}
\begin{tabular}{|c|c|c|c|c|}
\hline\rule{-2.5pt}{2.5ex}  
1 & 2 & 3 & 4 & 5\\
\hline
\end{tabular}
~ $\rightarrow$ ~
\begin{tabular}{|c|c|c|c|c|}
\hline\rule{-2.5pt}{2.5ex}
4 & 5 & 1 & 2 & 3\\
\hline
\end{tabular}
\end{center}
%\vspace{11pt}

Unlike the shift operator, the circular shift is invertible, i.e., there always exists a circular shift that will
return to the initial state. In fact, the operation can be expressed as an $n\times n$ 
permutation matrix of the following form:
\begin{equation}
P ~=~ 
\left[
\begin{array}{c|c}
  0 \cdots 0 & 1  \\ \hline
  \raisebox{-15pt}{{\huge\mbox{{I}}}} & ~ \\[-4.5ex]
  ~ & 0\\[-1.0ex]
  \raisebox{-5pt}{\mbox{\tiny (n-1\! $\times$\! n-1)}}  & \vdots \\[-0.6ex]
  ~ & 0
\end{array}
\right]
\end{equation}
where {\bf I} represents the identity matrix. In the cases of $n=3$ and $n=4$ this would give, 
respectively:
\begin{equation}
P_{\hspace{-2pt}\raisebox{-3pt}{\mbox{\tiny 3x3}}} = 
\left[
\begin{array}{ccc}
0 & 0 & 1\\
1 & 0 & 0\\
0 & 1 & 0
\end{array}\right] ~ \mbox{and} ~~
P_{\hspace{-2pt}\raisebox{-3pt}{\mbox{\tiny 4x4}}} = 
\left[
\begin{array}{cccc}
0 & 0 & 0 & 1\\
1 & 0 & 0 & 0\\
0 & 1 & 0 & 0\\
0 & 0 & 1 & 0
\end{array}\right]\!.
\end{equation}
A simple circular shift of the vector $[1~2~3]$ can thus be obtained by applying
$P_{\hspace{-2pt}\raisebox{-3pt}{\mbox{\tiny 3x3}}}$ as 
\begin{equation}
\left[
\begin{array}{ccc}
0 & 0 & 1\\
1 & 0 & 0\\
0 & 1 & 0
\end{array}\right] 
\cdot
\left[\begin{tabular}{c}
1 \\ 2 \\ 3 
\end{tabular}\right]
~ = ~
\left[\begin{tabular}{c}
3 \\ 1 \\ 2
\end{tabular}\right].
\end{equation}

Raising $P$ to an integral power $k$, i.e., $P^k$, has the effect of shifting a vector
by $k$ positions. In the case of $n=5$, for example, a circular shift of $k=2$ can be 
expressed as
\begin{equation}
\left[\begin{tabular}{ccccc}
0 & 0 & 0 & 0 & 1\\
1 & 0 & 0 & 0 & 0\\
0 & 1 & 0 & 0 & 0\\
0 & 0 & 1 & 0 & 0\\
0 & 0 & 0 & 1 & 0
\end{tabular}\right]^{\large 2} \cdot
\left[\begin{tabular}{c}
1 \\ 2 \\ 3 \\ 4 \\ 5
\end{tabular}\right] ~ = ~ 
\left[\begin{tabular}{c}
4 \\ 5 \\ 1 \\ 2 \\ 3
\end{tabular}\right]
\end{equation}
This provides a clear generalization from an integer to a real-valued parameter $k$ because
the transformation $P^k$ is well-defined for any real value for the exponent. 

A natural question to ask is whether a simpler interpolation-based generalization, e.g., one that
applies some sort of weighted average based on the fractional part of the parameter, might 
represent a superior alternative. The principal obstacle to
defining such a generalization is ensuring that it is invertible, i.e., maintains all information.
As an example, consider the vector $[1~3~1~3]$ with $k=1/2$. In this case almost any 
simple interpolation scheme will produce a value for each location that is the mean of the 
value shifted half-way into the location and the value shifted half-way out of the location, thus 
yielding the result $[2~2~2~2]$. The vector of all equal values is a fixed point for almost 
any generalization of the circular-shift operator, and any non-invertible scheme will tend toward
that fixed point at the expense of information about the original state\footnote{It can be verified that 
the GCS vector transform defined using powers of the $n$-dimensional circular-shift matrix has 
the all-equal state as an eigenvector such that it is invariant with respect to any choice of $k$.}.
In other words, performing a sequence of non-integer interpolation-type operations to rows
and/or columns of the 8-tile puzzle would lose information in a way that prevents it from
being solved, i.e., being able to reach the goal state.

Having identified a mathematically consistent generalization of the operator, a practical
concern arises about the efficiency with which it can be evaluated. Specifically, for a 
plausible-size value of $n$, say $n=5$, can a given non-integral power of a $5\times 5$
matrix be evaluated efficiently enough to satisfy real-time constraints? The answer of course
depends on the computational resources available. The general algorithm for raising a matrix
to an arbitrary real exponent is unlikely to take more than a fraction of a second on a
typical smart phone, so at worst there might be a slight noticeable lag after completion of
a swipe by the user because the display update cannot begin until after the transformation
has been completed. Fortunately, the GCS operator does not involve an arbitrary matrix.
A closer examination reveals that the circular-shift matrix is {\em circulant}, i.e., each
row $i$ is equal to row $i-1$ circularly shifted by one position. Circulant matrices are 
special (\cite{gray,golub}) in that they can be diagonalized in $O(n^2\log(n))$ time, 
as opposed to $O(n^3)$ for a general matrix, and consequently can
be raised to an arbitrary real power and multiplied by a given vector with the same
complexity\footnote{More specifically, because $P$ is circulant it is diagonalizable by 
the Fourier matrix, which can be applied using the fast discrete Fourier transform (DFT)
in $\Theta(n \log(n))$ time\,\cite{fft,fftcomp}. If a fixed matrix logarithm 
$L$ is selected, which 
is necessary to ensure solution uniqueness for non-integral powers, $P^k$ can be 
evaluated as $\exp(kL)$, which after diagonalization only requires the exponential to 
be applied to the product of $k$ and the vector of eigenvalues of $L$.}. The matrix $P$ is also 
special in that it is unitary\,\cite{hj1}, i.e., the Euclidean norm of $Pv$ will equal that of $v$, 
and it has equal row and column sums, which means that the sum the elements of $Pv$ will 
equal that of $v$. Taken jointly, these two properties can be summarized as preserving 
the mean and variance of the elements of the transformed vector.

In the following section we present our main result: an $n$-parameter nonlinear matrix
operator based on the GCS.

\section{GCS Matrix Transform Operators}

We now generalize the vector GCS operator of the previous section to a matrix 
operator, denoted $\RowCS$. It must be emphasized that the operator under
consideration is {\em not} of the form $P^k M$ or $M P^k$, where all rows
or columns undergo the same linear transformation. Rather, the GCS matrix
operator developed in this section performs separately-parameterized circular 
rotations of individual rows or columns. For example, in the case of a $3\times 3$ 
matrix the rows can be circularly rotated with three independent parameters 
$\alpha,\beta,\gamma$ as:
\begin{equation}
\left[\!\!\begin{tabular}{c}
$\alpha$ \\ $\beta$ \\ $\gamma$
\end{tabular}\!\!\right]
\RowCS
\left[\!\!\begin{tabular}{ccc}
a & b & c\\
d & e & f\\
g & h & i
\end{tabular}\!\!\right]
\doteq
\left[\!\!\begin{tabular}{c}
\mbox{GCS}$\,\left(\mbox{[a~b~c]},\,\alpha\right)$\\
\mbox{GCS}$\,\left(\mbox{[d~e~f]},\,\beta\right)$\\
\mbox{GCS}$\,\left(\mbox{[g~h~i]},\,\gamma\right)$
\end{tabular}\!\!\right]
\end{equation}
where the operator $\RowCS$ indicates that the $i$th
element of the parameter vector defines the circular shift
to be applied to the $i$th {\em row} of the matrix. 

A converse operator, denoted $\ColCS$, is analogously defined
to operate on the {\em columns} of the matrix as
\begin{equation}
\left[\!\!\begin{tabular}{c}
$\alpha$ \\ $\beta$ \\ $\gamma$
\end{tabular}\!\!\right]
\ColCS
\left[\!\!\begin{tabular}{ccc}
a & b & c\\
d & e & f\\
g & h & i
\end{tabular}\!\!\right]
\doteq
\left[\!\!\begin{tabular}{c}
\mbox{GCS}$\,\left(\mbox{[a~d~g]},\,\alpha\right)$\\
\mbox{GCS}$\,\left(\mbox{[b~e~h]},\,\beta\right)$\\
\mbox{GCS}$\,\left(\mbox{[c~f~\,i]},\,\gamma\right)$
\end{tabular}\!\!\right]^{\mbox{\small T}}
\end{equation}
where the first column of the resulting matrix
is {GCS}$\,\left(\mbox{[a~d~g]},\,\alpha\right)$.

If $\RowCS$ and $\ColCS$ are taken as right-associative then a sequence of applications
can be meaningfully interpreted, e.g.,
\begin{eqnarray}
\left[\!\!\begin{tabular}{c}
$\xi$ \\ $\psi$ \\ $\omega$
\end{tabular}\!\!\right]
\ColCS
\left[\!\!\begin{tabular}{c}
$\alpha$ \\ $\beta$ \\ $\gamma$
\end{tabular}\!\!\right]
\RowCS
\left[\!\!\begin{tabular}{ccc}
a & b & c\\
d & e & f\\
g & h & i
\end{tabular}\!\!\right]
~\equiv~\hspace*{0.75in}\nonumber\\
\hspace*{0.75in} \left[\!\!\begin{tabular}{c}
$\xi$ \\ $\psi$ \\ $\omega$
\end{tabular}\!\!\right]
\ColCS
\left(\left[\!\!\begin{tabular}{c}
$\alpha$ \\ $\beta$ \\ $\gamma$
\end{tabular}\!\!\right]
\RowCS
\left[\!\!\begin{tabular}{ccc}
a & b & c\\
d & e & f\\
g & h & i
\end{tabular}\!\!\right]\right)\!\!.
\end{eqnarray}

It can be verified that the inverse of either operator can be obtained by
taking the negative of its parameters:
\begin{equation}
\left[\!\!\!\!\begin{tabular}{c}
$-\alpha$ \\ $-\beta$ \\ $-\gamma$
\end{tabular}\!\!\right]
\RowCS
\left[\!\!\begin{tabular}{c}
$\alpha$ \\ $\beta$ \\ $\gamma$
\end{tabular}\!\!\right]
\RowCS
\left[\!\!\begin{tabular}{ccc}
a & b & c\\
d & e & f\\
g & h & i
\end{tabular}\!\!\right]
=
\left[\!\!\begin{tabular}{ccc}
a & b & c\\
d & e & f\\
g & h & i
\end{tabular}\!\!\right]
\end{equation}
and more generally that
\begin{eqnarray}
\left[\!\!\begin{tabular}{c}
$\alpha$ \\ $\beta$ \\ $\gamma$
\end{tabular}\!\!\right]
\RowCS
\left[\!\!\begin{tabular}{c}
$\xi$ \\ $\psi$ \\ $\omega$
\end{tabular}\!\!\right]
\RowCS
\left[\!\!\begin{tabular}{ccc}
a & b & c\\
d & e & f\\
g & h & i
\end{tabular}\!\!\right]
~=~ \hspace*{0.75in} \nonumber \\
\hspace*{0.75in} \left[\!\!\begin{tabular}{c}
$\alpha+\xi$ \\ $\beta+\psi$ \\ $\gamma+\omega$
\end{tabular}\!\!\right]
\RowCS
\left[\!\!\begin{tabular}{ccc}
a & b & c\\
d & e & f\\
g & h & i
\end{tabular}\!\!\right]
\end{eqnarray}
but that no similar composition is possible for a mixed
sequence of $\RowCS$ and $\ColCS$ operations, e.g.,
they do not commute:
\begin{eqnarray}
\left[\!\!\begin{tabular}{c}
$\alpha$ \\ $\beta$ \\ $\gamma$
\end{tabular}\!\!\right]
\RowCS
\left[\!\!\begin{tabular}{c}
$\phi$ \\ $\xi$ \\ $\psi$ \\ $\omega$
\end{tabular}\!\!\right]
\ColCS
\left[\!\!\begin{tabular}{cccc}
a & b & c & d\\
e & f & g & h\\
i & j & k & l
\end{tabular}\!\!\right]
~\neq~\hspace*{0.75in} \nonumber \\
\hspace*{0.75in}\left[\!\!\begin{tabular}{c}
$\phi$ \\ $\xi$ \\ $\psi$ \\ $\omega$
\end{tabular}\!\!\right]
\ColCS
\left[\!\!\begin{tabular}{c}
$\alpha$ \\ $\beta$ \\ $\gamma$
\end{tabular}\!\!\right]
\RowCS
\left[\!\!\begin{tabular}{cccc}
a & b & c & d\\
e & f & g & h\\
i & j & k & l
\end{tabular}\!\!\right]
\end{eqnarray}
and, as shown above, do not have the same number of parameters
if the matrix argument is not square.

The line (i.e., row or column) operations performed by the $\RowCS$ 
and $\ColCS$ operators are separately parameterized and thus do not
generally represent a linear transformation. However, they preserve key 
structural properties such as the norm and sum of each line. In other 
words, while they can dramatically affect the determinant and other 
scalar functions of a given matrix, they do so without scaling any
row or column. A simple though illustrative example is the following
\begin{equation}
\left[\!\!\begin{tabular}{c}
0 \\ -1 \\ -2
\end{tabular}\!\!\right]
\RowCS
\left[\!\!\begin{tabular}{ccc}
1 & 0 & 0\\
0 & 1 & 0\\
0 & 0 & 1
\end{tabular}\!\!\right]
~=~
\left[\!\!\begin{tabular}{ccc}
1 & 0 & 0\\
1 & 0 & 0\\
1 & 0 & 0
\end{tabular}\!\!\right]
\end{equation}
where the result of the transformation is a reduction
of a full-rank matrix to a matrix of rank 1. The following
example shows that the same degree of rank reduction
is also possible for a full-rank matrix even when all of its values 
are distinct and no two rows or columns have the same 
norm or sum:
\begin{equation}
\left[\!\!\!\begin{tabular}{c}
1.2 \\ 1.4 \\ 1.8
\end{tabular}\!\!\!\right]
\!\!\RowCS\!\!
\left[\!\!\begin{tabular}{ccc}
3.1484\!\! & 1.3213\!\! & 1.5303\\
9.2946\!\! & 5.2798\!\! & 3.4257\\
4.9393\!\! & 5.3574\!\! & 1.7033
\end{tabular}\!\!\right]
\!\!\!=\!\!
\left[\!\!\begin{tabular}{ccc}
1\!\! & 2\!\! & 3\\
3\!\! & 6\!\! & 9\\
2\!\! & 4\!\! & 6
\end{tabular}\!\!\right]\!\!.
\label{rank}
\end{equation}
Of course this example was contrived to ensure a rank-1 result,
but it nonetheless demonstrates that strong shift and scale relationships
among the rows may be difficult to discern\footnote{It can be hypothesized
that spatial coherence of elements within matrices representing, e.g., natural 
images, may tend to create such relationships, at least locally. If so then 
GCS-based methods may provide a basis for defining new types of matrix
decompositions/splittings \cite{hj1,mil} and for approximate rank reduction 
that is distinct from conventional approaches. Again, this is a topic beyond
the scope of the present paper.}. The principal motive for
considering rank properties is that they can be exploited by an automated
$n$-tile puzzle solver whenever the goal matrix is not of full rank, e.g., as
would be true in cases such as:
\begin{equation}
\left[\!\!\begin{tabular}{ccc}
1 & 2 & 3\\
4 & 5 & 6\\
7 & 8 & 9
\end{tabular}\!\!\right] \,\mbox{or}~
\left[\!\!\begin{tabular}{ccc}
0 & 1 & 0\\
1 & 1 & 1\\
0 & 1 & 0
\end{tabular}\!\!\right] \,\mbox{or}~
\left[\!\!\begin{tabular}{ccc}
1 & 0 & 1\\
0 & 1 & 0\\
1 & 0 & 1
\end{tabular}\!\!\right]\!\! .
\end{equation}
However, our main focus in this paper is on interactive game-play
involving a human user, which is developed in the next section.

To summarize, the operators defined in this section intended for
use in representing circular shifts of rows and columns of a matrix as
performed when solving a sliding-puzzle game. For example, suppose
the final two moves of the game involve a circular shift of the 
third row by 2 positions, followed by a 1-position shift of the middle
column, then the sequence can be expressed using the new
operators as
\begin{equation}
\left[\!\!\begin{tabular}{c}
0 \\ 0 \\ 2
\end{tabular}\!\!\right]
\RowCS
\left[\!\!\begin{tabular}{ccc}
1 & 5 & 3\\
4 & 8 & 6\\
9 & 7 & 2
\end{tabular}\!\!\right]
=
\left[\!\!\begin{tabular}{ccc}
1 & 5 & 3\\
4 & 8 & 6\\
7 & 2 & 9
\end{tabular}\!\!\right]
\end{equation}
followed by
\begin{equation}
\left[\!\!\begin{tabular}{c}
0 \\ 1 \\ 0
\end{tabular}\!\!\right]
\ColCS
\left[\!\!\begin{tabular}{ccc}
1 & 5 & 3\\
4 & 8 & 6\\
7 & 2 & 9
\end{tabular}\!\!\right]
~=~
\left[\!\!\begin{tabular}{ccc}
1 & 2 & 3\\
4 & 5 & 6\\
7 & 8 & 9
\end{tabular}\!\!\right]
\end{equation}
or the consecutive operations can
be expressed jointly as
\begin{equation}
\left[\!\!\begin{tabular}{c}
0 \\ 1 \\ 0
\end{tabular}\!\!\right]
\ColCS
\left[\!\!\begin{tabular}{c}
0 \\ 0 \\ 2
\end{tabular}\!\!\right]
\RowCS
\left[\!\!\begin{tabular}{ccc}
1 & 5 & 3\\
4 & 8 & 6\\
9 & 7 & 2
\end{tabular}\!\!\right]
=
\left[\!\!\begin{tabular}{ccc}
1 & 2 & 3\\
4 & 5 & 6\\
7 & 8 & 9
\end{tabular}\!\!\right]\!\!.
\end{equation}

\section{The GCS 9-Tile Puzzle}

As suggested by the example of Eq.\,\ref{rank}, the application of
real-valued shifts will transform tiles with integer values to ones with
non-integer values, e.g., 
\begin{equation}
\left[\!\!\!\begin{tabular}{c}
1.08 \\ 0.61 \\ 2.64
\end{tabular}\!\!\!\right]
\!\RowCS\!
\left[\!\!\!\begin{tabular}{ccc}
1\!\! & 2\!\! & 3\\
4\!\! & 5\!\! & 6\\
7\!\! & 8\!\! & 9
\end{tabular}\!\!\!\right]
\!\!=\!\!
\left[\!\!\!\begin{tabular}{ccc}
 3.0823\!\!\! &  1.1103\!\!\! &  1.8074\\
 5.2637\!\!\! &  3.8946\!\!\! &  5.8417\\
 6.8758\!\!\! &  8.7904\!\!\! &  8.3337
\end{tabular}\!\!\!\right]\!\!.
\label{puzzle1}
\end{equation}
This of course should be expected, but now it is necessary to 
seriously consider whether such a generalization is conducive to 
engaging gameplay. The original version of the
game had numbers on the tiles, but from a player's perspective they are 
just symbols to be moved. By contrast, the generalized form above displays 
an inherently mathematical aspect that most people are unlikely to find 
accessible, let alone enjoyable. Assuming the focus is not on entertaining
our future robot overlords, there is a clear need to abstract away from
raw numeric values and redefine in terms of something with more direct
intuitive appeal. This can be accomplished by treating the numeric values
as {\em parameters} for determining what is displayed on the tiles. For
example, the numeric values could be mapped to grey-scale colors to
to give the following alternative display for the original goal configuration:
\vspace{-8pt}\begin{center}
\includegraphics[width=0.475\textwidth]{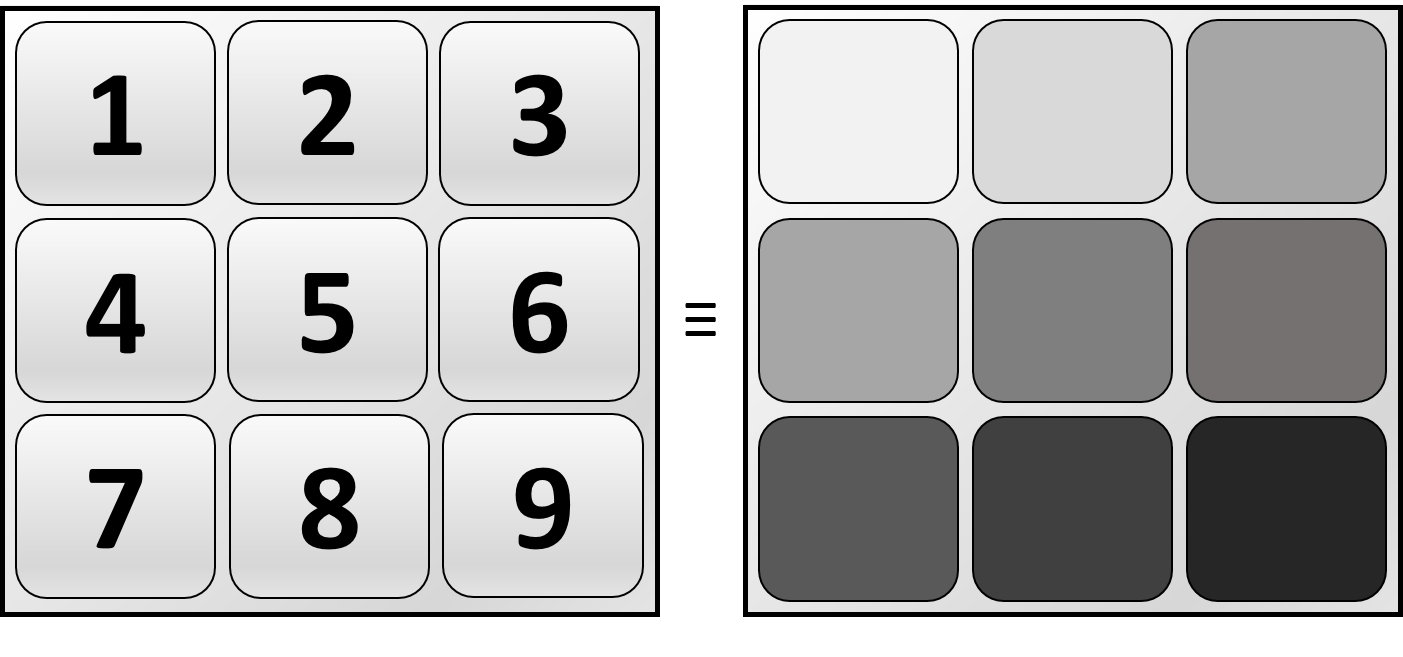}
\end{center}
\vspace{-12pt}
This is encouraging, but whereas the increasing sequence $1...9$
represents a distinctive goal configuration in the case of tiles with numeric
digits, the sequence of increasingly darker shades of grey is not so
distinctive because humans have difficulty distinguishing {\em absolute},
as opposed to {\em relative}, gradations of color intensity. In other 
words, a player may have difficulty assessing whether a given
sequence of shades of grey {\em exactly} equals the desired
sequence for any given row or column. What is needed is a 
completely unambiguous goal state. In this case the ideal goal
state would be the uniform state in which all tiles have the same 
color.

Unfortunately, the GCS operation has no effect on a vector of identical 
values, so there is no way to initialize in the uniform state and 
``scramble'' to a non-uniform state by applying a sequence of
random circular shifts to the rows and columns. That's not necessarily
an issue, though,
because the functional mapping of numeric values to grey-scale values 
can be changed to whatever we please to satisfy our needs. For example, 
we can define the numeric goal state to be 
\begin{equation}
\left[\begin{tabular}{rrr}
-1 & 1 & 1\\
1 & -1 & 1\\
1 & 1 & -1
\end{tabular}\right]
\end{equation}
and use their absolute values (unsigned magnitudes) to determine
their grey-scale intensities. This will create a multiplicity of matrices
that correspond to the goal state, e.g.,
\begin{equation}
\left[\begin{tabular}{rrr}
1 & 1 & -1\\
-1 & 1 & 1\\
1 & -1 & 1
\end{tabular}\right]
\end{equation}
but that can easily be accommodated by only checking absolute
values when assessing whether the player has achieved a/the 
goal state. To summarize, the player will begin with a 
configuration of tiles having a random distribution of 
grey-scale intensities and will have to perform a sequence
of swipes to circularly shift the rows and columns until
the tiles have uniform color:
\vspace{-8pt}\begin{center}
\includegraphics[width=0.475\textwidth]{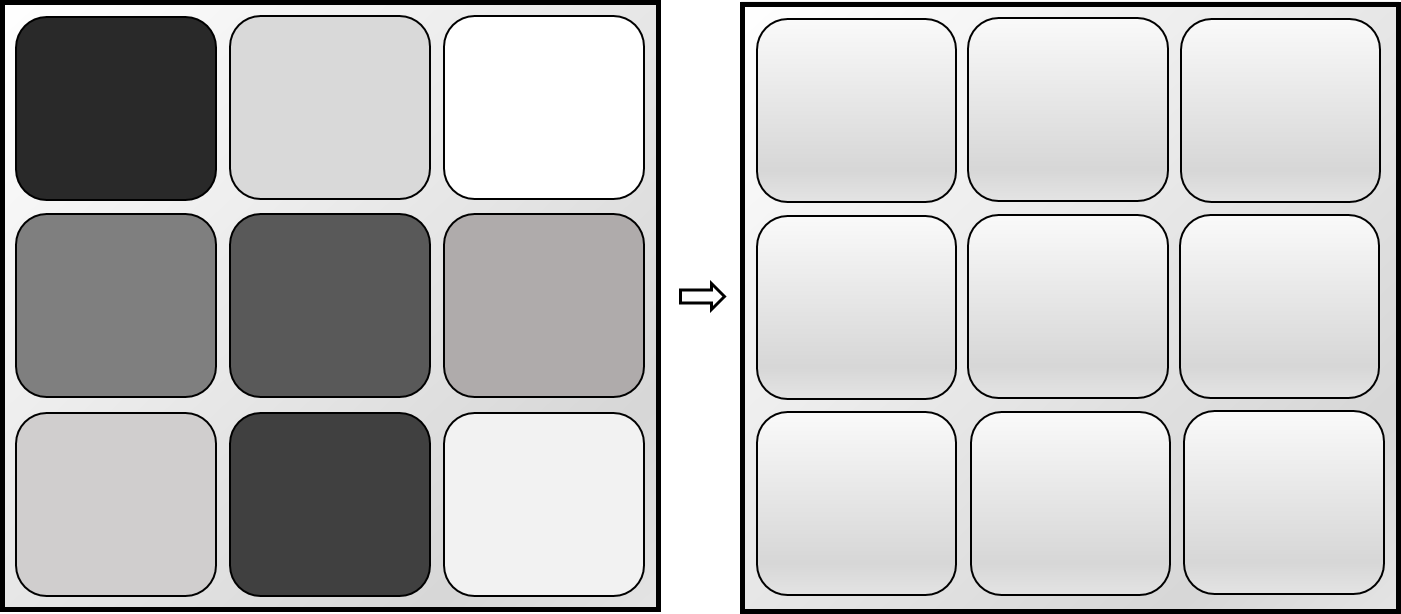}
\end{center}
\vspace{-8pt}
It should be noted that unlike the discrete values of the original
game, the values in the generalized version are not suitable
for the use of equality tests. This means that a threshold
should be applied so that, e.g., if all values are within $0.05$
of having unit magnitude then the user should be deemed 
to have achieved the goal state.

\section{Further Generalizations}

The most obvious generalization would be to expand from a 
$3\times 3$ matrix of tiles to a $4\times 4$ matrix. However,
an important fact that has not been mentioned about the 
generalized circular-shift operator is that it yields complex
numbers in even dimensions, e.g.:
\begin{eqnarray}
\left[\!\!\!\!\begin{tabular}{r}
1.3  \\ 3.2 \\ -2.2 \\ 2.8
\end{tabular}\!\!\right]
\!\RowCS\!
\left[\!\!\begin{tabular}{c}
2.7 \\ 3.2 \\ 2.4 \\ 1.2
\end{tabular}\!\!\right]
\!\ColCS\!
\left[\!\!\!\begin{tabular}{rrrr}
1 & 2 & 3 & 4\\
5 & 6 & 7 & 8\\
9 & 10 & 11 & 12\\
13 & 14 & 15 & 16
\end{tabular}\!\!\right]
= \hspace*{1.75in} \nonumber \\
\left[\!\!\!\!\begin{tabular}{rrrr}
14.89+1.63i\!\!\!  & 11.08-2.15i\!\!\!  & 2.83+1.76i\!\!\!  & 8.13-2.40i\\
 9.67-2.65i\!\!\!  & 14.17+3.32i\!\!\!  &  7.93-2.56i\!\!\!  & 9.40+3.06i\\
 5.63-1.05i\!\!\!  &  7.83+0.38i\!\!\!  & 10.43-0.78i\!\!\!  & 14.77+0.29i\\
 3.19-2.36i\!\!\!  &  6.40+3.13i\!\!\!  &  8.87-2.45i\!\!\!  & 0.78+2.86i
\end{tabular}\!\!\!\right]\!\!.\hspace*{1.0in} \nonumber
\end{eqnarray}
Implementing a game in this case using a
display of complex numeric values could pose a risk to humanity 
by potentially frustrating and provoking our future robot overlords,
but we have seen that this may be avoided by transforming 
to a non-numeric display. In fact, the same absolute-value 
approach used in the previous section can be applied identically
to complex numbers that arise in the $4\times 4$ case. 
Alternatively, we could also exploit the angle and magnitude
information given by complex numbers to generate a richer
variety of colors or to display directional intensity gradients
on the tiles.

Another possible generalization is to associate a matrix with
each tile instead of a single numeric value. This is essentially
what is done in the case of traditional puzzles in which the sliding 
tiles represent patches of an image. The following is 
an example of a $3\times 3$ puzzle in which $8$ virtual tiles 
depict parts of an image and one space is empty to permit
movement of the tiles. The image on the right is the goal state 
while the image on the left is the initial state obtained by jumbling 
the $8$ tiles (and implicitly the hole) using a random sequence 
of ordinary tile shifts:
\vspace{-8pt}\begin{center}
\includegraphics[width=0.475\textwidth]{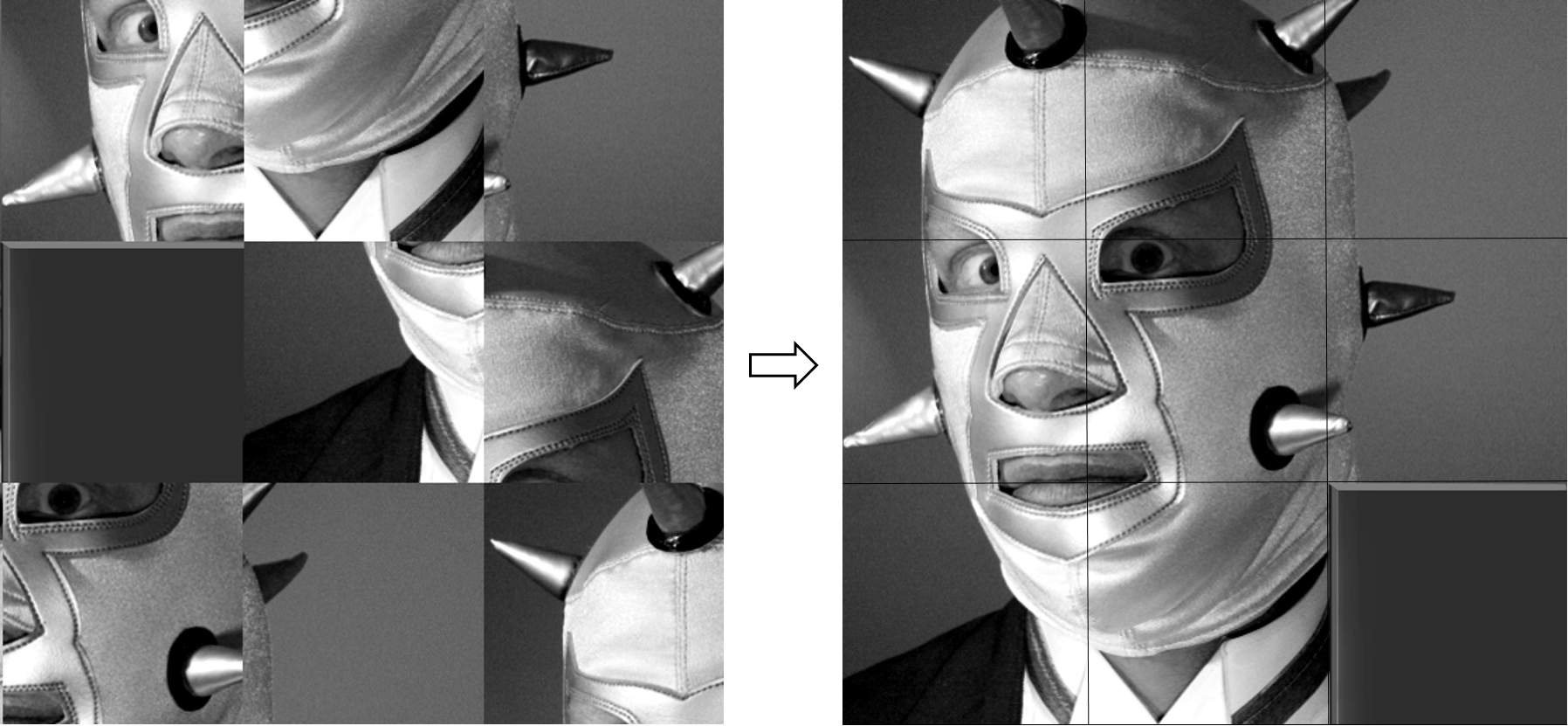}
\end{center}
\vspace{-8pt}  
The generalization of the circular-shift matrix $P$ for the shifting
of submatrices associated with tiles (e.g., to eliminate the need 
for an empty space/hole) 
can be achieved using the Kronecker matrix product
\begin{equation}
\mbox{I}\, \otimes P
\end{equation}
where $\mbox{I}$ is the identity matrix of size
equal to the size of the tiles to be shifted. Basically, this just
replaces each $i,j$ element of $P$ with $P_{i,j}$ times the 
identity matrix of size equal
to the block size. The following is the generalization of the
previous example where the full image can be represented 
on $9$ tiles because generalized circular shift operations
eliminate need for an empty space. Analogous to the
previous example, the image on the left is the initial state
obtained by jumbling the $9$ tiles using a random sequence
of generalized circular shifts:
\vspace{-8pt} \begin{center}
\includegraphics[width=0.475\textwidth]{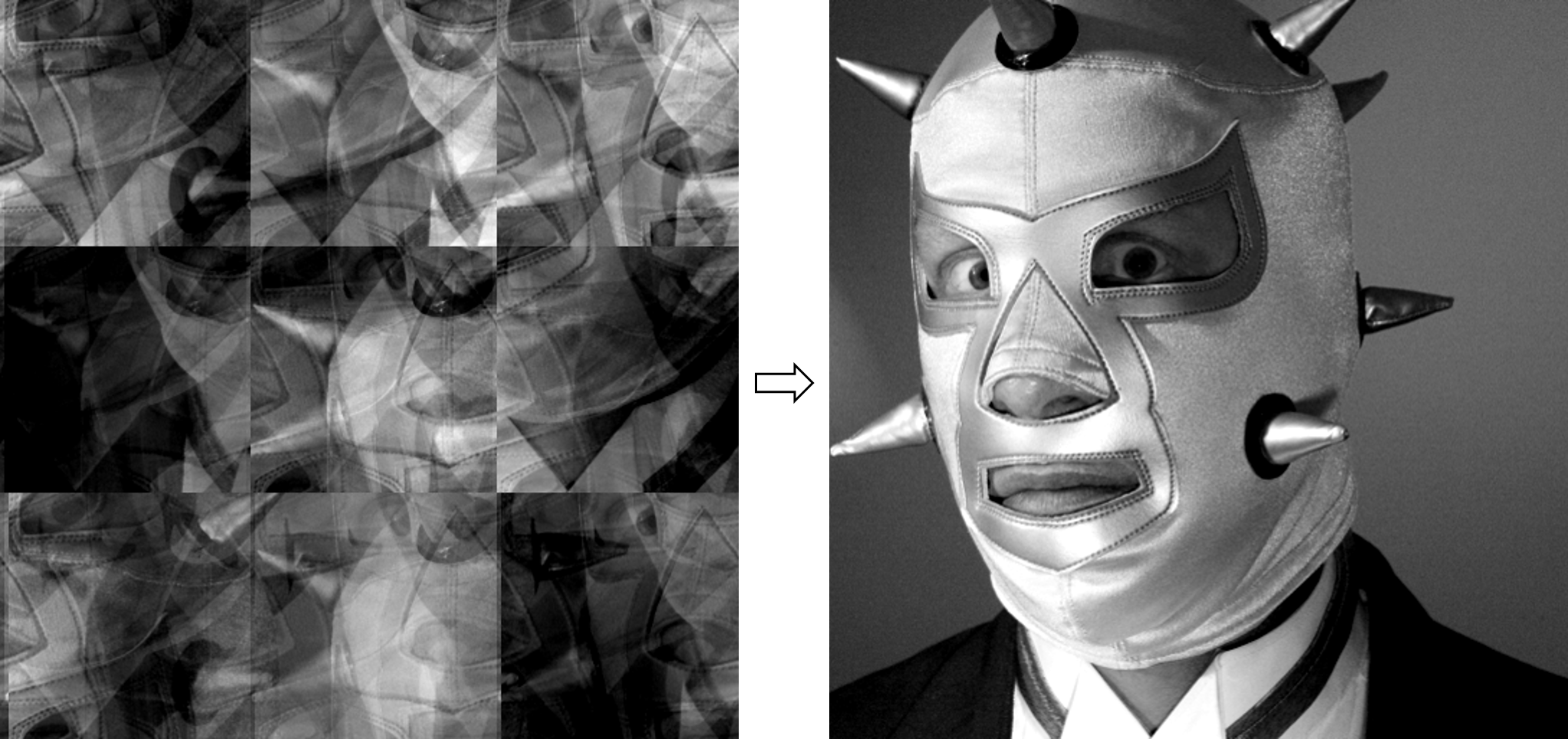}
\end{center}
\vspace{-8pt}  
As can be seen, the underlying matrix for the sub-image 
associated with each tile will generally be a superposition
(mixture) of 
the tile matrices in its row and column, so the extent to which
players can develop an intuitive {\em feel} for the effect
that a swipe has on a given row or column will impact
whether or not the puzzle is entertaining to solve. In
the $3\times 3$ case (e.g., like the above example)
the puzzle is not difficult to solve and may very well
prove to be more entertaining than simply moving static tiles because the final image
is less obvious at the outset and thus may produce a
greater degree of satisfaction/reward
when it finally clicks into place. As has been
mentioned, however, assessment of the quality of the
game requires user studies\,\cite{humfac} and is not 
the focus of this paper. As exemplified by the following 
image, the general aesthetic qualities of the GCS
transformation may be of independent interest\,\cite{joig}:
\vspace{-8pt}\begin{center}
\includegraphics[width=0.295\textwidth]{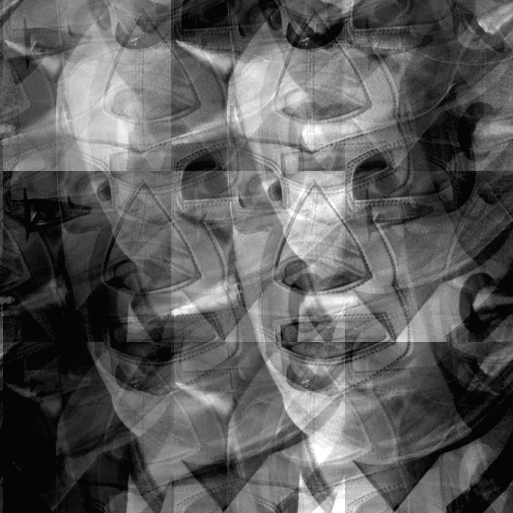}
\end{center}
\vspace{-8pt}  

\section{Discussion}
\vspace*{-4pt}
In this paper we have considered some of the opportunities 
available when adapting physical games and puzzles 
for computer simulation. In particular, we examined how
linear algebra, and efficient methods for performing certain 
matrix operations, can be applied to develop new operators
for generalizing traditional sliding-tile puzzles. Although our
principal focus has not been on any particular theory of
game design\,\cite{gamedesign1,gamedesign2}, the exercise has shown
how general considerations can lead to interesting engineering
challenges and potential solutions. The principal takeaway is
that the potential value of a strong engineering and mathematical
background for the design and adaptation of games is often
underestimated.

\end{document}